\documentclass[aps,prd,fleqn,superscriptaddress]{revtex4}
\pdfoutput=1
\usepackage{graphicx,color,natbib}
\usepackage{amsmath,amssymb,amsfonts}
\usepackage{epsfig,epsf}
\usepackage{dcolumn}
\usepackage{float}
\usepackage{bm}
\input amssym.def
\input amssym.tex

\makeatletter
\newcommand{\vast}{\bBigg@{3}}
\newcommand{\Vast}{\bBigg@{4}}
\newcommand{\bse}{\begin{subequations}}
\newcommand{\ese}{\end{subequations}}
\newcommand{\be}{\begin{equation}}
\newcommand{\ee}{\end{equation}}
\newcommand{\bea}{\begin{eqnarray}}
\newcommand{\eea}{\end{eqnarray}}
\newcommand{\ba}{\begin{array}}
\newcommand{\ea}{\end{array}}
\newcommand{\h}{\frac{1}{2}}

\def\m{\frac{\mu}{T}}

\usepackage[colorlinks=true, linkcolor=blue, bookmarks=true]{hyperref}

\begin{document}
\title{Temperature Dependence of Entanglement of Purification \\
in Presence of Chemical Potential}
\author{B. Amrahi\footnote{$\rm{b}_{-}$amrahi@sbu.ac.ir}}
\affiliation{Department of Physics, Shahid Beheshti University, Tehran, Iran}
\author{M. Ali-Akbari\footnote{$\rm{m}_{-}$aliakbari@sbu.ac.ir}}
\affiliation{Department of Physics, Shahid Beheshti University, Tehran, Iran}
\author{M. Asadi\footnote{$\rm{m}_{-}$asadi@ipm.ir}}
\affiliation{IPM, School of Particles and Accelerators, P.O. Box 19395-5531, Tehran, Iran}

\begin{abstract}
Using holographic idea, we study the entanglement of purification in a field theory with a critical point in intermediate and low temperature regime. This theory includes temperature $T$ as well as chemical potential $\mu$.
In the intermediate regime, due to chemical potential, we observe that new terms proportional to temperature square appear in the final result of entanglement of purification or equivalently, apart from $T^0$ and $T^4$ terms in the case of $\mu=0$, it contains the terms proportional to $T^2$.
Our results also indicate that the entanglement of purification, i.e. the correlation between subsystems, can decrease or increase depending on the value of $\m$  when the other parameters are kept fixed. However, in the low temperature limit, the correlation always decreases, comparing to the $\mu=0$ case, independent of the value of $\m$ when the other parameters do not alter. The existence of a critical point in the theory changes the behavior of entanglement of purification in such a way that the entanglement of purification experiences a maximum or minimum. Moreover, near the critical point, we analytically show that the critical exponent is equal to 0.5 in both regimes and also the term proportional to $T^2$ changes sign and becomes negative in the intermediate regime.
\end{abstract}

\maketitle

\tableofcontents

\section{Introduction}
How to calculate the dependence (correlation) between two subsystems has always been an important question. In the case of {\it{pure}} state, described by density matrix $\rho$, entanglement entropy, as a non-local observable in quantum field theory, measures the quantum correlation between subsystem $A$ and its complementary, $\bar{A}$. The entanglement entropy of subsystem $A$ is then defined as 
\be %
S_A=-{\rm{Tr}}(\rho_A\log\rho_A),
\ee
where $S_A$ states how much entanglement exists between subsystems. The reduced density matrix $\rho_A={\rm{Tr}}_{\bar{A}} (\rho)$ is obtained by tracing over the degrees of freedom in the complementary subsystem. In  the case of {\it{mixed state}}, described by density matrix $\rho_{AB\equiv A\cup B}$, the entanglement of purification (EoP) between two subsystems $A$ and $B$ measures the total (classical as well as quantum) correlation \cite{Bhattacharyya:2019tsi,  arXiv:quant-ph/0202044v3}. In fact, the EoP is a measure of correlation in terms of the entanglement of pure state. The EoP is then defined as follows: consider a purification state $|\psi\rangle_{ABA'B'}$ of a mixed state $\rho_{AB}$, that is a pure state in an enlarged Hilbert space ${\cal{H}}_A\otimes{\cal{H}}_B\otimes{\cal{H}}_{A'}\otimes{\cal{H}}_{B'}$ with the following constraint 
\be 
\rho_{AB}={\rm{Tr}}_{A'B'}\big[\left(|\psi\rangle_{ABA'B'}\right) \left({}_{ABA'B'}\langle\psi|\right)\big].
\ee 
Hence, the EoP is obtained by the minimal entanglement entropy over all purifications $|\psi\rangle_{ABA'B'}$, i.e. 
\be\label{eof}
E_p(\rho_{AB})=\underset{|\psi\rangle_{ABA'B'}}{\rm{min}} S_{AA'},
\ee
where $S_{AA'}$ is the entanglement entropy corresponding to $\rho_{AA'}$ and $\rho_{AA'}={\rm{Tr}}_{BB'}\big[\left(|\psi\rangle_{ABA'B'}\right) \left({}_{ABA'B'}\langle\psi|\right)\big]$. This definition for the EoP reduces to the entanglement entropy of subsystem $A$ if one considers the subsystem $B$ as complementary of subsystem $A$.

Gauge/gravity duality provides a remarkable connection between a classical gravity in $d + 1$ dimensions and a certain strongly coupled gauge theory in $d$ dimensions \cite{Maldacena:1997re}. This duality has been applied to study different areas of physics such as quantum chromodynamics phase transition, various properties of quark-gluon plasma and condensed matter \cite{CasalderreySolana:2011us,Hartnoll:2009sz}. The gauge/gravity duality generally proposes an applicable and simple geometrical interpretation on the gravity side for a complicated phenomenon in the strongly coupled gauge theory. As an example, Ryu and Takayanagi firstly proposed in \cite{Ryu:2006bv} that the entanglement entropy can be computed from
\be 
S_A =\frac{{\rm{Area}}(\Gamma_A)}{4G_N},
\ee
where $G_N$ is the $d$-dimensional Newton constant. $\Gamma_A$ is a codimension-2 minimal surface, called RT-surface, whose boundary $\partial_A \Gamma$
coincides with the boundary of the subregion $A$ on the boundary of the gravity theory where, according to the gauge/gravity dictionary the quantum field theory lives, i.e. $\partial_A \Gamma=\partial A$. In fact, based on this duality in order to calculate the entanglement entropy in a field theory we only need to compute an extremal surface in gravity dual.

\begin{figure}
\includegraphics[width=100 mm]{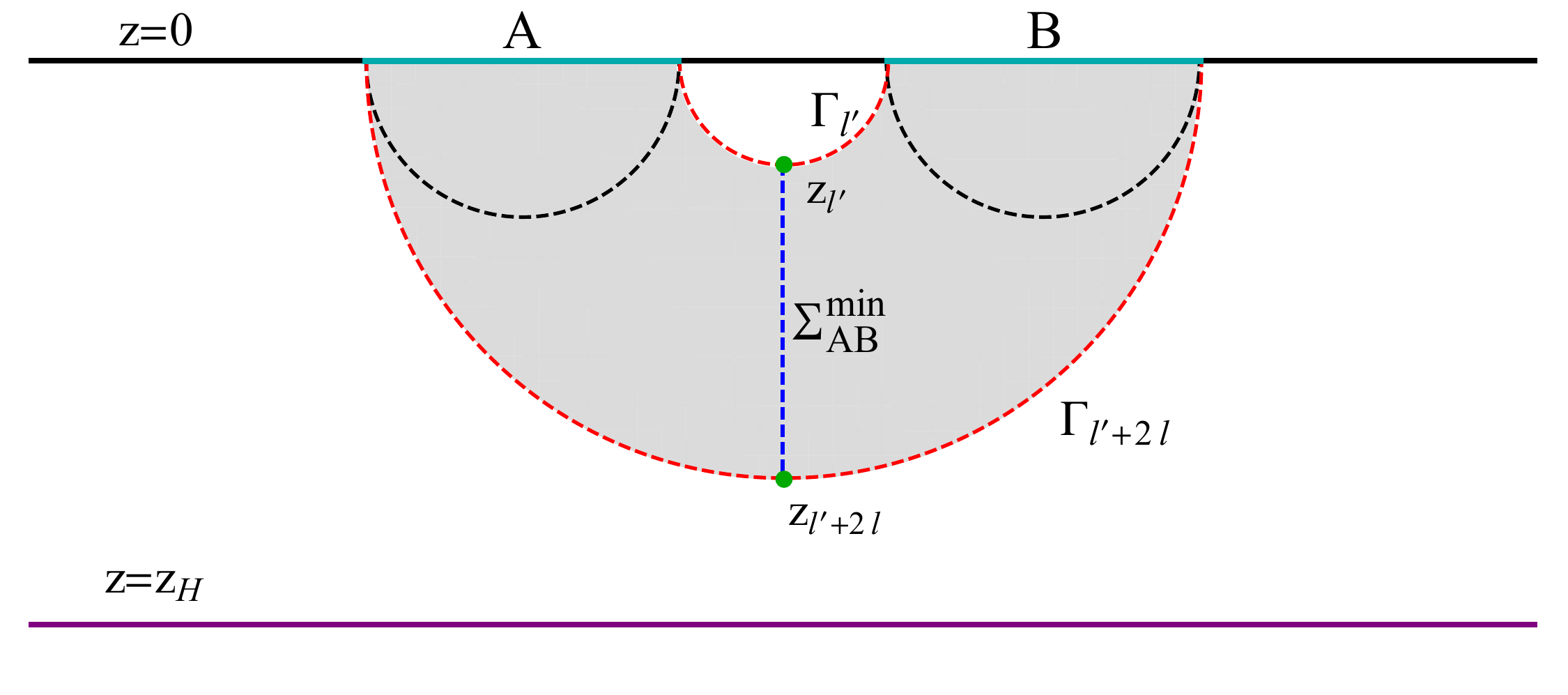}
\caption{The gray region shows the entanglement wedge dual to $\rho_{AB}$. The minimal surfaces, RT-surfaces, are denoted by $\Gamma$, the dashed curves.}
\label{fig2}
\end{figure}

As an another example, in the context of information theory, a new quantity which recently received a lot of interest in the gauge/gravity duality point of view is EoP. As a matter of fact, it has been conjectured that the EoP is holographically dual to entanglement wedge cross-section $E_w$ of $\rho_{AB}$, as a measure of total correlation between $A$ and $B$, which is defined \cite{Takayanagi:2017knl, Nguyen:2017yqw}
\be\label{eop} %
E_w(\rho_{AB})=\frac{{\rm{Area}}(\Sigma_{AB}^{min})}{4G_N},
\ee
where $\Sigma_{AB}^{min}$ is the minimal area surface in the entanglement wedge $E_w(\rho_{AB})$ that ends on the RT-surface of $A\cup B$, the blue-dashed line in figure \ref{fig2}.
As a result, we have
\be \label{eop1}
E_p(\rho_{AB})\equiv E_w(\rho_{AB}).
\ee
It is also discussed in the literature that the entanglement wedge cross-section $E_w$ of $\rho_{AB}$ may be identified with logarithmic negativity \cite{Kudler-Flam:2018qjo}, odd entropy \cite{Tamaoka:2018ned}, entanglement distillation \cite{Agon:2018lwq} and reflected entropy \cite{Dutta:2019gen}. For more details, we refer the interested reader to the original literature.
% The (logarithmic) negativity is a computable measure for quantum mixed states derived from the positive partial transpose criterion for the separability of mixed states and is defined by taking the trace norm of the partial transpose of the density matrix \cite{Kudler-Flam:2018qjo}. The odd entanglement entropy is a generalization of the entanglement entropy that measures the correlation between two subsystems in a mixed state. This quantity can be calculated by using the replica trick and enables us to compute the entanglement wedge cross-section in holographic CFTs \cite{Tamaoka:2018ned}. The EoP of mixed states and therefore the entanglement wedge cross-section can be interpreted in terms of entanglement distillation \cite{Agon:2018lwq}. The reflected entropy is defined for mixed quantum states and is the entanglement entropy associated with a canonical purification which is obtained by doubling the Hilbert space \cite{Dutta:2019gen}.

In this paper, we are going to study the effect of chemical potential (and temperature) on the EoP analytically. In other words, the gravity solution, which is dual to the our state in the field theory, has charge as well as mass. Moreover, since this theory enjoys a critical point we are able to investigate the behavior of EoP near the critical point and obtain critical exponent.
\section{Review on the holographic background}
Since we would like to study a (3+1)-dimensional filed theory with a critical point holographically, we start with the following 5-dimensional action \cite{Gubser:1998jb}
\be
S=\frac{1}{16 \pi G_N^{(5)}}\int d^5x\sqrt{-g}\left({\cal{R}}-\frac{f(\phi)}{4}F_{\mu\nu}F^{\mu\nu}-\h(\partial_\mu\phi)^2-V(\phi)\right),
\ee
where $G_N^{(5)}$ is the 5-dimensional Newton constant. $g$ and ${\cal{R}}$ are the determinant of the metric and its corresponding Ricci scalar, respectively. $F_{\mu\nu}$ is field strength of the gauge field and $\phi$ denotes dilaton field. The dilaton potential, $V(\phi)$, and the Maxwell-Dilaton coupling, $f(\phi)$, are given by
\be
V(\phi)=-\frac{1}{R^2}(8e^{\frac{\phi}{\sqrt{6}}}+4 e^{-\sqrt{\frac{2}{3}}\phi}) ,\ \ \ \ \ \ f(\phi)= e^{-2\sqrt{\frac{2}{3}}\phi}.
\ee
Solving the equations of motion, one can find a solution, known as the 1RCBH background, and it describes by the following metric
\begin{align}\label{1RCBH}
ds^2=e^{2A(z)}\bigg(-h(z)dt^2+d\vec{x}^2\bigg)+\frac{e^{2B(z)}}{h(z)}\frac{R^4}{z^4}dz^2,
\end{align}
where
\begin{align}\label{coef}
 \begin{split}
A(z)&=\ln\left(\frac{R}{z}\right)+\frac{1}{6}\ln \left(1+\frac{Q^2z^2}{R^4}\right),\cr 
B(z)&=-\ln\left(\frac{R}{z}\right)-\frac{1}{3}\ln \left(1+\frac{Q^2z^2}{R^4}\right),\cr
h(z)&=1-\frac{M^2z^4}{R^6(1+\frac{Q^2z^2}{R^4})}. \cr
\phi(z)&=-\sqrt{\frac{2}{3}} \ln\left(1+\frac{Q^2z^2}{R^4}\right),\cr
A_t(z)&=\frac{MQz_H^2}{R^4\left(1+\frac{Q^2z_H^2}{R^4}\right)}-\frac{MQz^2}{R^4\left(1+\frac{Q^2z^2}{R^4}\right)},
\end{split}
\end{align}
where $A_t(z)$ is the time component of the gauge field which has been chosen to be zero on the horizon and regular on the boundary. The above metric describes a charged black hole background. $M$ and $Q$ are the black hole mass and charge, respectively and $R$ is the $AdS$ radius supposed to be one hereafter. The 3+1-dimensional spacetime, which the dual field theory lives on, denotes by $(t, \vec{x}=(x_1,x_2,x_3))$ and $z$ is the radial bulk coordinate and the strongly coupled field theory lives on the boundary at $z \to 0$. The location of the black hole horizon is obtained from $h(z_H)=0$ as follows
\begin{align}
z_H=\sqrt{\frac{Q^2+\sqrt{Q^4+4M^2}}{2M^2}}.
\end{align}
The temperature $T$ and chemical potential $\mu$ of dual field theory are given by
\begin{align}\label{mut}
\begin{split}
T&=\frac{1}{2\pi z_H}\left(\frac{2+Q^2z_H^2}{\sqrt{1+Q^2z_H^2}}\right),\cr 
\mu &=\frac{Q}{\sqrt{1+Q^2z_H^2}}.
\end{split}
\end{align}
The field theory, which is dual to metric \eqref{1RCBH} in the limit of zero charge, is clearly conformal. One therefore expects that all physical quantities can be written as a function of dimensionless parameter $\mu/T$. It is then shown that the phase diagram of the dual field theory is simple and can be described by a one-dimensional line which ends at a critical point, i.e. $\mu/T=\pi/\sqrt{2}$ \cite{DeWolfe:2011ts,Finazzo:2016psx}. In other words, using \eqref{mut}, one easily finds
 \begin{align}\label{QzH}
 \xi^\frac{1}{2}\equiv Qz_H=\sqrt{2}\left(\frac{1\pm \sqrt{1-\left(\frac{\mu/T}{\pi/\sqrt{2}}\right)^2}}{\frac{\mu/T}{\pi/\sqrt{2}}}\right).
 \end{align}
Since $Qz_H$ is real and nonnegative, the above equation indicates that $\mu /T\in\left[0,\pi/\sqrt{2}\right]$. As a result, for a given value of $\mu/T$, there are obviously two distinct values for $Qz_H$, which parametrize two different branches of solutions. The thermodynamics of metric \eqref{1RCBH} has been extensively discussed, for instance see \cite{DeWolfe:2011ts,Finazzo:2016psx,Ebrahim:2020qif}, and it is shown that the branch with $Q z_H< \sqrt{2}\ (\mu/T<\pi/\sqrt{2})$ is stable one or in other words we choose the minus sign in \eqref{QzH}.

\section{Entanglement of purification}
In the previous section, we reviewed the background which is dual to the strongly coupled field theory with a critical point. Investigating properties of various physical observables near the critical point is an attractive question. Observables such as entanglement entropy, mutual information, and quasinormal mode have been already discussed \cite{Ebrahim:2020qif,Finazzo:2016psx} and in the following, we are interested in studying the EoP near a critical point. Luckily, similar to the case of entanglement entropy, EoP has also a simple holographic counterpart \cite{Takayanagi:2017knl, Nguyen:2017yqw} and we in fact need to calculate a minimal area (divided by 4$G_N^{(5)}$) which is defined as the entanglement wedge cross section of a mixed state. 

In order to compute the EoP holographically, we use background \eqref{1RCBH} and consider a symmetric configuration of two strips at a given time slice as subsystems $A$ and $B$, i.e. two parallel strips with equal widths $l$ extended along $x_2$ and $x_3$ directions with length $L(\rightarrow \infty)$ and separated by a distance $l'$, let's say in the $x_1$ direction, see figure \ref{fig2}. In this figure, one clearly sees that $\Sigma_{AB}^{min}$ runs along the radial direction and connect the turning points of minimal surfaces $\Gamma_{l'}$ and $\Gamma_{l'+2l}$. Applying the holographic prescription introduced in \cite{Takayanagi:2017knl, Nguyen:2017yqw}, the area of this hypersurface turns out to be
\begin{align}\label{EoP1}
E_w=\frac{L^2}{4G_N^{(5)}}\int_{z^*_{l'}}^{z^*_{l'+2l}}dz \frac{e^{2A(z)+B(z)}}{z^2\sqrt{h(z)}},
\end{align}
where $z^*_{l'}$ and $z^*_{l'+2l}$ denote the turning point of $\Gamma_{l'}$ and $\Gamma_{l'+2l}$, respectively. To find $z^*_{l'}$, we consider the following configuration
\begin{align}
x_1(z)\equiv x(z)\in [-\frac{l'}{2}, \frac{l'}{2}] \ , \ \ \ \ \ \ \ 
x_i\in[-\frac{L}{2}, \frac{L}{2}] \ ,\ \ \ \ \ i=2,3 ,
\end{align}
and by calculating the area of this configuration, we find 
\begin{align}\label{area}
Area=2L^2\int_0^{z^*_{l'}} dz e^{3A(z)}\sqrt{x'(z)^2+\frac{e^{2B(z)-2A(z)}}{z^4 h(z)}}.
\end{align}
Since there is no explicit $x(z)$ dependence in \eqref{area}, the corresponding Hamiltonian is constant and one can easily obtain
\begin{align}\label{turning}
\frac{l'}{2}=\int_0^{z^*_{l'}}dz \frac{e^{B(z)-A(z)}}{z^2h(z)}\frac{e^{3A(z^*_{l'})}}{\sqrt{e^{6A(z)}-e^{A(z^*_{l'})}}},
\end{align}
where the constant is chosen to be $\frac{e^{B(z)-A(z)}}{z^2\sqrt{h(z)}}\sqrt{e^{6A(z)}-e^{6A(z^*_{l'})}}$. Similar equation for $z^*_{l'+2l}$ is obtained by replacing $l$ and $z^*_{l'}$ with $l'+2l$ and $z^*_{l'+2l}$ in above equation, respectively. Substituting the explicit form of the metric components \eqref{coef} in  \eqref{EoP1} and \eqref{turning}, we have
\begin{align}\label{EoP2}
E_p\equiv 4G_N^{(5)} E_w=L^2\int_{z^*_{l'}}^{z^*_{l'+2l}}dz \frac{1}{z^3}\sqrt{1+\xi(\frac{z}{z_H})^2}\left(1+\xi(\frac{z}{z_H})^2-(1+\xi)(\frac{z}{z_H})^4\right)^{-\frac{1}{2}},
\end{align}
\begin{align}\label{length}
\frac{l'}{2}=\int_0^{z^*_{l'}}dz \left(1+\xi(\frac{z}{z_H})^2-(1+\xi)(\frac{z}{z_H})^4\right)^{-\frac{1}{2}}(\frac{z}{z^*_{l'}})^3\left(\frac{1+\xi(\frac{z^*_{l'}}{z_H})^2}{1+\xi(\frac{z}{z_H})^2}\right)^\frac{1}{2}\left[1-(\frac{z}{z^*_{l'}})^6\left(\frac{1+\xi(\frac{z^*_{l'}}{z_H})^2}{1+\xi(\frac{z}{z_H})^2}\right)\right]^{-\frac{1}{2}}.
\end{align}
In order to calculate the above integrals, we need to use the following (binomial and trinomial) expansions
\bse\begin{align}
\label{expansion1} (1+x)^r&=\sum\limits_{n=0}^{\infty}\binom{r}{n}x^n ,\ \ \ \ \ \ \ |x|<1,\\
\label{expansion2} (1+x)^{-r}&=\sum\limits_{n=0}^{\infty}(-1)^n\binom{r+n-1}{n}x^n , \ \ \ \ \ \ \ |x|<1, \\ 
%\label{expansion3} (1+x+y)^r&=\sum\limits_{k=0}^{\infty}\sum\limits_{n=0}^{k} \binom{r}{k}\binom{k}{n}x^{k-n}y^n,\ \ \ \ \ \ |x+y|<1,\\ 
\label{expansion4} (1+x+y)^{-r}&=\sum\limits_{k=0}^{\infty}\sum\limits_{n=0}^{k} (-1)^k\binom{r+k-1}{k}\binom{k}{n}x^{k-n}y^n , \ \ \ \ \ \ \  |x+y|<1,
\end{align}
\ese
where $x$, $y$ and $r$ are real numbers and $r>0$. Using \eqref{expansion1} by identifying $x=\xi (\frac{z}{z_H})^2$ and \eqref{expansion4} by identifying $x=\xi (\frac{z}{z_H})^2$ and $y=-(1+\xi)(\frac{z}{z_H})^4$ and  finally by taking the integrals, it is straightforward to obtain
\begin{align}\label{finalEoP}
\begin{split}
E_{p}=\frac{L^2}{2}\sum\limits_{k=0}^{\infty}\sum\limits_{n=0}^{k}\sum\limits_{m=0}^{\infty}&\frac{\Gamma(k+\frac{1}{2})}{\Gamma(n+1)\Gamma(k-n+1)\Gamma(m+1)\Gamma(\frac{3}{2}-m)}\cr
&\times\frac{(-1)^{k+n}\xi^{k-n+m}(1+\xi)^n}{2(n+k+m-1)}\frac{{z^*_{l'+2l}}^{2(n+k+m-1)}-{z^*_{l'}}^{2(n+k+m-1)}}{z_H^{2(n+k+m)}},
\end{split}
\end{align}
where the condition $n+k+m\neq 1$ must be satisfied. Moreover, one can check that $|x|<1$ and $|x+y|<1$ for any allowable values of background parameters, that is $\xi \in [0,2]$ and $z\ll z_H$. Similarly, identifying $x=\xi (\frac{z}{z_H})^2$, $y=-(1+\xi)(\frac{z}{z_H})^4$ and using \eqref{expansion4} and also identifying $x=(\frac{z}{z^*_{l'}})^6\left(\frac{1+\xi(\frac{z^*_{l'}}{z_H})^2}{1+\xi(\frac{z}{z_H})^2}\right)$ and using \eqref{expansion2} and finally by taking the integrals, equation \eqref{length} leads to 
\begin{align}\label{Length1}
\begin{split}
\frac{l'}{2}=z^*_{l'}\sum\limits_{k=0}^{\infty}\sum\limits_{n=0}^{k}\sum\limits_{m=0}^{\infty}\sum\limits_{j=0}^{\infty}&\frac{\Gamma(k+\frac{1}{2})\Gamma(j+m+\frac{1}{2})\Gamma(2+3j+k+n)}{\Gamma(n+1)\Gamma(k-n+1)\Gamma(3+3j+k+n+m)\Gamma(j+1)}\cr 
&\times(-1)^{k+n}\xi^{k-n+m}(1+\xi)^n\left[1+\xi(\frac{z^*_{l'}}{z})^2\right]^{-m}(\frac{z^*_{l'}}{z_H})^{2(m+n+k)},
\end{split}
\end{align}
and then it is easy to check that the conditions required to use the binomial and trinomial expansions for negative powers are satisfied as follows  
\begin{align}
\left(\frac{z}{z^*}\right)^6\left(\frac{1+\xi \left(\frac{z^*}{z_H}\right)^2}{1+\xi \left(\frac{z}{z_H}\right)^2}\right)<1 \ \ \ \ \ \&  \ \ \ \ \frac{\xi \left(\frac{z^*}{z_H}\right)^2}{1+\xi \left(\frac{z^*}{z_H}\right)^2}\left(1-\frac{z^2}{{z^*}^2}\right)<1 .
\end{align}
Now, we have to solve \eqref{Length1} for $z^*_{l'}$ and use it in \eqref{finalEoP} to get the EoP in terms of $l'$. Unfortunately, it is not possible to solve \eqref{Length1} analytically, and hence, from now on, we will mainly consider only some orders of expansion in different regimes such as low ($Tl\ \&\ Tl'\ll 1$) and intermediate ($T l' \ll 1\ll T l$) temperature regimes.
\subsection{Low temperature regime}
In the limit of low temperature, i.e. $T l\ \&\ T l'\ll 1$ or equivalently ${z^*}_{l'}\ \&\ {z^*}_{l'+2l}\ll z_H$, the extremal surfaces $\Gamma_{l'}$ and $\Gamma_{l'+2l}$ are restricted to be near the boundary and therefore the leading contribution to the EoP comes from the $AdS$ boundary. Finite temperature corrections, in the presence of chemical potential, appear as sub-leading terms corresponding to the deviation of the bulk geometry from pure $AdS$ and we will compute them perturbatively. Thus, we expand \eqref{Length1} up to 4th order in $\frac{z^*_{l'}}{z_H}$ and finally have
\begin{align}\label{length4}
l'=z^*_{l'}\left\lbrace a_1-\frac{a_1\xi}{6}(\frac{z^*_{l'}}{z_H})^2+\left[\frac{a_2(1+\xi)}{2}+\frac{a_3\xi^2}{24}\right](\frac{z^*_{l'}}{z_H})^4\right\rbrace+\mathcal{O}(\frac{z^*_{l'}}{z_H})^6,
\end{align}
where $a_1$, $a_2$ and $a_3$ are given by
\begin{align}
\begin{split}
a_1&=\frac{3\sqrt{\pi}\Gamma(\frac{5}{3})}{\Gamma(\frac{1}{6})},\ \ \ \ \ \ \ 
a_2=\frac{\sqrt{\pi}\Gamma(\frac{7}{3})}{4\Gamma(\frac{11}{6})},\cr
a_3&=\frac{3}{\sqrt{\pi}}\left[\Gamma(\frac{5}{6})\Gamma(\frac{5}{3})-\frac{3}{5}\Gamma(\frac{7}{6})\Gamma(\frac{7}{3})\right]-\frac{1}{70}  {}_{3}F_{2}\left(\frac{3}{2},\frac{5}{3},\frac{7}{3};\frac{8}{3},\frac{10}{3};1\right).
\end{split}
\end{align}
Solving \eqref{length4} for $z^*_{l'}$ perturbatively up to 4th order in $\frac{l'}{z_H}$, we obtain
\begin{align}\label{z in low}
z^*_{l'}=\frac{l}{a_1}\left\lbrace 1+\frac{\xi}{6a_1^2}(\frac{l'}{z_H})^2+\frac{1}{2a_1^4}\left[\frac{\xi^2}{6}(1-\frac{a_3}{2a_1})-\frac{a_2}{a_1}(1+\xi)\right](\frac{l'}{z_H})^4
\right\rbrace +\mathcal{O}(\frac{l'}{z_H})^6.
\end{align}
In the same way, one can find an expression for ${z^*}_{l'+2l}$. It is easy to see that its functionality is the same as \eqref{z in low} and one only needs to substitute $z^*_{l'}$ and $l'$ with ${z^*}_{l'+2l}$ and $l'+2l$, respectively.
Replacing $z^*_{l'}$ and  ${z^*}_{l'+2l}$ in \eqref{finalEoP}, we obtain
\begin{align}\label{ahmad}
E_p=\frac{L^2a_1^2}{2}\left\lbrace\frac{1}{l'^2}-\frac{1}{(l'+2l)^2}\right\rbrace - \frac{2L^2\pi^4}{a_1^2}\left[\frac{\xi^2}{12}(\frac{a_3}{a_1}-1)+(\frac{a_2}{a_1}-\frac{1}{2})(1+\xi)\right]
\left[\frac{(1+\xi)^2}{(1+\frac{\xi}{2})^4}\right]l(l'+l)T^4+{...} ,
\end{align}
where $\xi$ in terms of $\mu/T$ has been introduced in \eqref{QzH} and as a result $E_p$ is a complicated function of $\mu/T$. We would like to emphasize a few points about our result:
\begin{figure}
\includegraphics[width=80 mm]{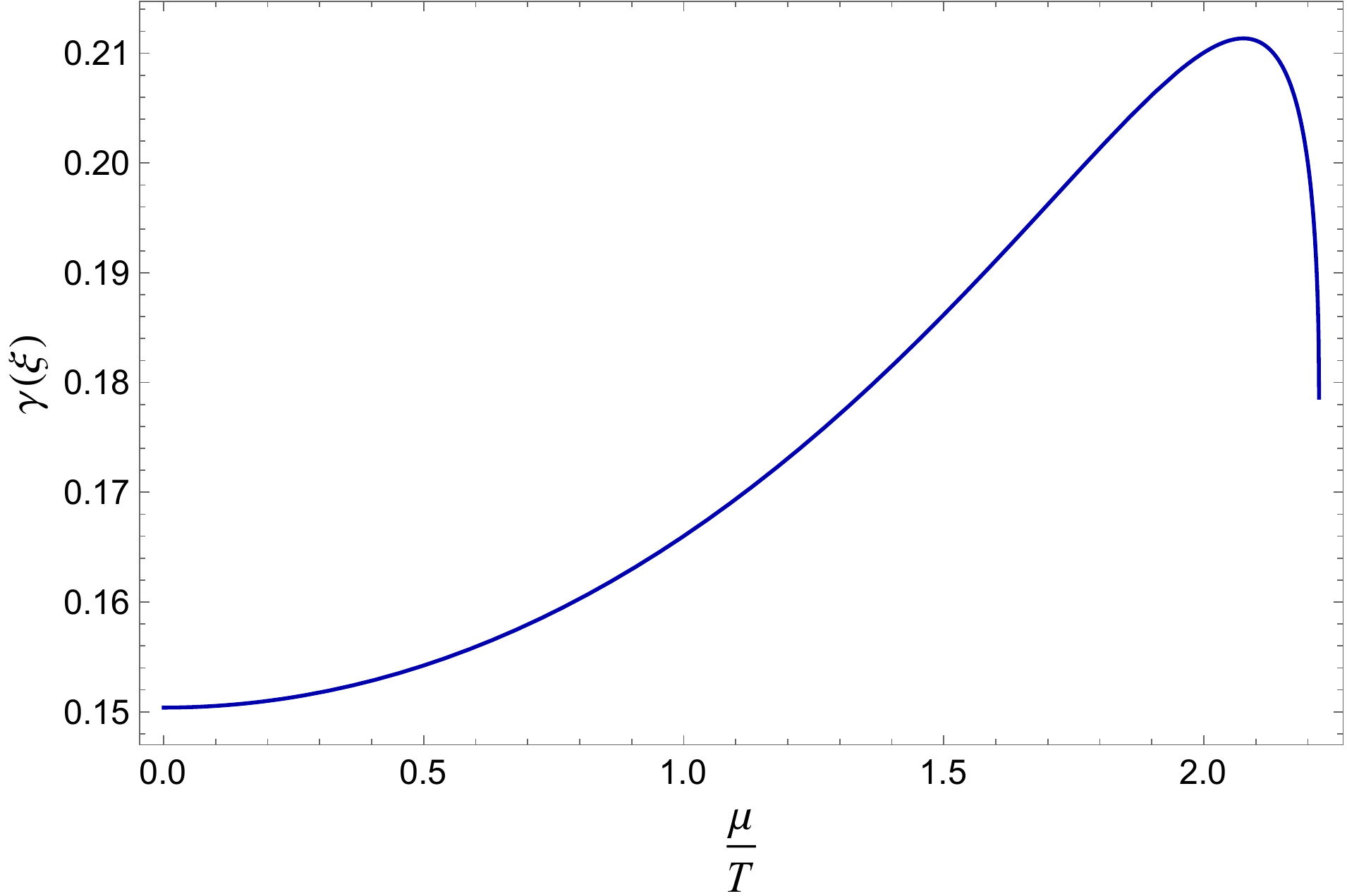}
\caption{$\gamma(\xi)$ in terms of $\m$.}
\label{correction}
\end{figure}
\begin{itemize}
\item The first term in the above equation is the leading term corresponding to the EoP of our configuration in zero temperature field theory. Obviously, this term is positive.
\item The terms which are proportional to $\xi^k,\ k>0$ are coming from the effect of non-zero chemical potential in the field theory. Note also that the constant term $a_3$ only appears in this case.
\item The term $\frac{a_2}{a_1}-\frac{1}{2}>0$ is always positive in the reliable regime of parameters and therefore temperature, for $\xi=0$, decreases the correlation between our considered subsystems. Obviously, the higher temperature, the bigger correction.
\item The term $\gamma(\xi)=\left(\frac{\xi^2}{12}(\frac{a_3}{a_1}-1)+(\frac{a_2}{a_1}-\frac{1}{2})\xi\right)\frac{(1+\xi)^2}{(1+\frac{\xi}{2})^4}>0$ is always positive and similar to the case of finite temperature, non-zero chemical potential decreases the correlation. We plot function $\gamma(\xi)$ in terms of $\m$ in figure \ref{correction}. It is clearly seen that this correction depends on the value of $\m$. Moreover, it shows that for arbitrary value of temperature one may find two distinct states, labeled with two different values of $\m$, with the same value of $\gamma(\xi) T^4$. In the case of zero chemical potential, there is one-to-one correspondence between temperature and its related correction since it is proportional to $T^4$. One may speculate the mentioned behavior happens because of existence of critical point. 
\item In the limit of $\xi=0$, we reproduce the previous results, see \cite{BabaeiVelni:2019pkw}. It seems that the minimum value of correction term happens at $\xi =0$.
\end{itemize}

\subsection{Intermediate temperature regime}
In this subsection, we study another interesting limit called intermediate regime. The limit is defined by $T l' \ll 1\ll T l$ or equivalently ${z^*}_{l'}\ll z_H$ and ${z^*}_{l'+2l}\to z_H$. In fact, the extremal surface $\Gamma_{l'}$ is restricted to be near the boundary and, on the contrary, the turning point of the extremal surface $\Gamma_{2l+l'}$ approaches the horizon, i.e. ${z^*}_{l'+2l}\to z_H$. We define $u=\frac{z}{z_H}$ and rewrite \eqref{EoP2}  as follows
\begin{align}\label{EoPP}
 E_p=\frac{L^2}{z_H^2}\int_{u_{l'}}^{u_{l'+2l}}du \frac{\sqrt{1+\xi u^2}}{u^3\sqrt{1-u^2}\sqrt{1+(1+\xi)u^2}},
\end{align}
where $u_{l'}=\frac{z^*_{l'}}{z_H}$ and $u_{l'+2l}=\frac{z^*_{l'+2l}}{z_H}$. Therefore, the intermediate temperature limit is equivalent to $u_{l'}\ll 1$ and $u_{l'+2l}\to 1$.  Using \eqref{EoPP} and \eqref{expansion2} by identifying $x=-u^2$, we get
\begin{align}
\begin{split}\label{EoPP1}
 E_p &=\frac{L^2}{z_H^2}\sum\limits_{n=0}^{\infty}\frac{\Gamma(n+\frac{1}{2})}{\sqrt{\pi}\Gamma(n+1)}\int_{u_{l'}}^{u_{l'+2l}}du \frac{u^{2n-3}\sqrt{1+\xi u^2}}{\sqrt{1+(1+\xi)u^2}}\cr
 &=\frac{L^2}{2z_H^2}\int_{u_{l'}}^{u_{l'+2l}}du \frac{\left(2u^{-3}-u^{-1}\right)\sqrt{1+\xi u^2}}{\sqrt{1+(1+\xi)u^2}}+\frac{L^2}{\sqrt{\pi} z_H^2}\sum\limits_{n=2}^{\infty}\frac{\Gamma(n+\frac{1}{2})}{\Gamma(n+1)}\int_{u_{l'}}^{u_{l'+2l}}du \frac{u^{2n-3}\sqrt{1+\xi u^2}}{\sqrt{1+(1+\xi)u^2}}
 \end{split}
\end{align}
Fortunately, we are able to take the above integrals analytically and it is thus easy to find 
\begin{align}
\begin{split}\label{EoPP2}
E_p&=\frac{L^2}{2z_H^2}\Bigg\lbrace -\frac{\sqrt{1+\xi u_{l'+2l}^2}\sqrt{1+(1+\xi)u_{l'+2l}^2}}{ u_{l'+2l}^2}+\frac{\xi u_{l'+2l}^2}{2}F_1\left(1;\frac{1}{2},\frac{1}{2};2;-\xi u_{l'+2l}^2,-(1+\xi)u_{l'+2l}^2\right)\cr
&+\sum\limits_{n=2}^{\infty}\frac{\Gamma(n+\frac{1}{2})}{\sqrt{\pi}(n-1)\Gamma(n+1)}u_{l'+2l}^{2n-2}F_1\left(n-1;-\frac{1}{2},\frac{1}{2};n;-\xi u_{l'+2l}^2,-(1+\xi)u_{l'+2l}^2\right)\cr
& +\frac{\sqrt{1+\xi u_{l'}^2}\sqrt{1+(1+\xi)u_{l'}^2}}{ u_{l'}^2}-\frac{\xi u_{l'}^2}{2}F_1\left(1;\frac{1}{2},\frac{1}{2};2;-\xi u_{l'}^2,-(1+\xi)u_{l'}^2\right)\cr
&-\sum\limits_{n=2}^{\infty}\frac{\Gamma(n+\frac{1}{2})}{\sqrt{\pi}(n-1)\Gamma(n+1)}u_{l'}^{2n-2}F_1\left(n-1;-\frac{1}{2},\frac{1}{2};n;-\xi u_{l'}^2,-(1+\xi)u_{l'}^2\right) \Bigg\rbrace ,
\end{split}
\end{align}
where $F_1(a;b_1,b_2;c;x,y)$ is the Appell hypergeometric function of two variables. For $\vert x\vert<1$ and $\vert y\vert<1$, the function $F_1$ is defined by the double series
\begin{align}
F_1(a;b_1,b_2;c;x,y)=\sum\limits_{n=0}^{\infty}\sum\limits_{m=0}^{\infty}\frac{(a)_{m+n}(b_1)_m(b_2)_n}{(c)_{m+n}m!n!}x^my^n,
\end{align}
where $(x)_n=\frac{\Gamma(x+1)}{\Gamma(x-n+1)}$  is the Pochhammer symbol. For other values of $x$ and $y$, the function $F_1$ is given by the following series
\begin{align}
F_1(a;b_1,b_2;c;x,y)=\sum\limits_{r=0}^{\infty}\frac{(a)_r(b_1)_r(b_2)_r (c-a)_r}{(c+r-1)_r (c)_{2r}r!} x^ry^r {}_{2}F_{1}\left(a+r,b_1+r;c+2r;x\right) {}_{2}F_{1}\left(a+r,b_2+r;c+2r;y\right),
\end{align}
where ${}_{2}F_{1}$ is the hypergeometric function. We should now check the convergence of the above series.  For large $n$, the infinite series in the second line in \eqref{EoPP2} goes as
\begin{align}
\frac{u_{l'+2l}^{2n-2}F_1\left(n;-\frac{1}{2},\frac{1}{2};n;-\xi u_{l'+2l}^2,-(1+\xi)u_{l'+2l}^2\right)}{n^\frac{3}{2}}=\frac{u_{l'+2l}^{2n-2}}{n^\frac{3}{2}}\frac{\sqrt{1+\xi u_{l'+2l}^2}}{\sqrt{1+(1+\xi)u_{l'+2l}^2}}.
\end{align}
In the limit of $u_{l'+2l}\to 1$, this term behaves as $n^{-\frac{3}{2}}$. Therefore, the series is convergent and we can safely take the mentioned limit. For another part of the intermediate regime, i.e. $u_{l'}\ll 1$, we expand \eqref{EoPP2} up to second order in $u_{l'}$ and using \eqref{z in low} for $z^*_{l'}$, one gets
\begin{align}
\begin{split}\label{EoPP3}
E_p=\frac{L^2}{2}\vast\lbrace &\frac{a_1^2}{l'^2}+\Bigg[\sum\limits_{n=2}^{\infty}\frac{\Gamma(n+\frac{1}{2})}{\sqrt{\pi}(n-1)\Gamma(n+1)}F_1\left(n-1;-\frac{1}{2},\frac{1}{2};n;-\xi ,-(1+\xi)\right)+\frac{1}{2}-\sqrt{(1+\xi)(2+\xi)}\cr
&+\frac{\xi}{2}\left(\frac{4}{3}+F_1\left(1;\frac{1}{2},\frac{1}{2};2;-\xi ,-(1+\xi)\right)\right)\Bigg]\frac{(1+\xi)\pi^2}{(1+\frac{\xi}{2})^2}T^2 \cr
&+\frac{\pi^4}{a_1^2}\left[\frac{\xi^2}{12}\left(\frac{a_3}{a_1}-1\right)+\left(\frac{a_2}{a_1}-\frac{1}{2}\right)(1+\xi)\right]\frac{(1+\xi)^2}{(1+\frac{\xi}{2})^4}l'^2T^4\vast\rbrace .
\end{split}
\end{align}
First of all, $l$ does not play any role in our final result. It seems reasonable since in numerical analysis \cite{Amrahi:2020jqg} it was shown that for large enough $l$, compared to $T^{-1}$, the EoP is independent of the length of subsystems. Furthermore, our result shows that in the limit of zero chemical potential, i.e. $\xi=0$, the terms which are proportional to $T^2$ disappear. Thus, due to non-zero $\mu$, a new contribution proportional to temperature square becomes manifest and we call them {\it{non-perturbative}} corrections since the temperature does not need to be small. However, the last term is a {\it{perturbative}} effect since by assumption $l'T$ is small enough. Another important point is that in the presence of chemical potential the correlation between two subsystems is sensitive to the value of $l'$ and temperature and not $l'T$.
Evidently, in the limit of $\xi=0$, we reproduce the results obtained for the thermal case, see \cite{BabaeiVelni:2019pkw}.

\begin{figure}
\includegraphics[width=80 mm]{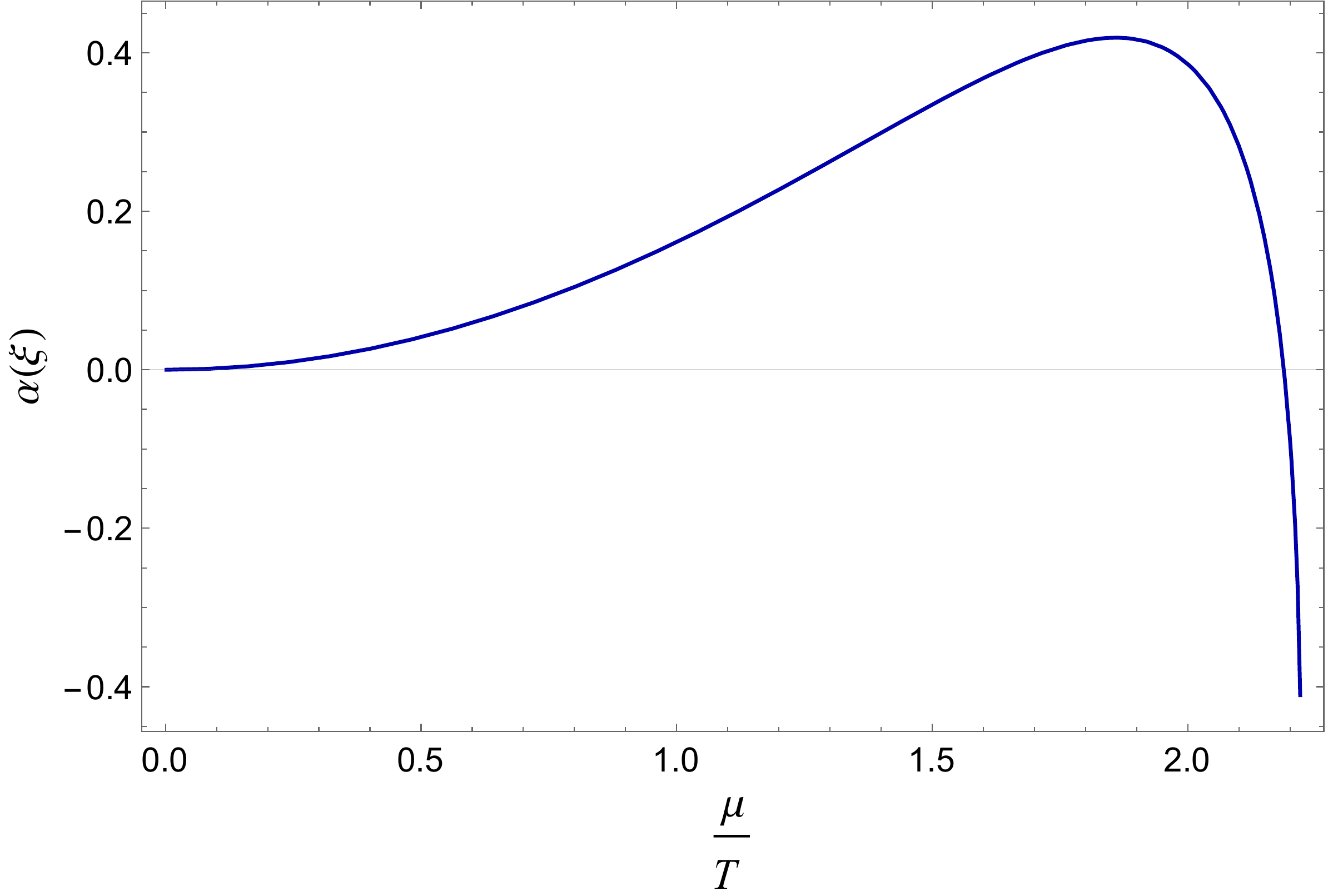}
\caption{$\alpha(\xi)$ in terms of $\frac{\mu}{T}$. The infinity in \eqref{EoPP3} is considered to be $n=40000$.}
\label{coeff}
\end{figure}

The last term in \eqref{EoPP3}, perturbative term, is positive. Then the terms which are proportional to $T^2$ is written as $\alpha(\xi)T^2$, where 
\begin{align}
\begin{split}
\alpha(\xi)&=\Bigg[\sum\limits_{n=2}^{\infty}\frac{\Gamma(n+\frac{1}{2})}{\sqrt{\pi}(n-1)\Gamma(n+1)}F_1\left(n-1;-\frac{1}{2},\frac{1}{2};n;-\xi ,-(1+\xi)\right)+\frac{1}{2}-\sqrt{(1+\xi)(2+\xi)}\cr
&+\frac{\xi}{2}\left(\frac{4}{3}+F_1\left(1;\frac{1}{2},\frac{1}{2};2;-\xi ,-(1+\xi)\right)\right)\Bigg]\frac{(1+\xi)\pi^2}{(1+\frac{\xi}{2})^2}.
\end{split}
\end{align}
In figure \ref{coeff}, we plot $\alpha(\xi)$ and this figure indicates that $\alpha(\xi)$ is finite and can be  positive or negative. As a result, in the limits we are considering here the term proportional to $l'^{-1}$  in \eqref{EoPP3} is dominant and temperature, depending on the value of chemical potential, increases or decreases the correlation between two subsystems.

\section{Critical exponent}
As it was mentioned, the field theory we are investigating in this paper enjoys a critical point at $\frac{\mu}{T}^*=\frac{\pi}{\sqrt{2}}$. The behavior of different observables near the critical point has been studied in the literature \cite{DeWolfe:2011ts, Finazzo:2016psx, Ebrahim:2017gvk, Ebrahim:2018uky, Ebrahim:2020qif,Amrahi:2020jqg} and it reveals that this behavior can be considered as $(\frac{\mu}{T}^*-\frac{\mu}{T})^{-\theta}$ where $\theta$ is the dynamical exponent and it is obtained to be 0.5. As matter of fact, it was shown that various quantities, such as entanglement entropy, mutual information, equilibration time and relaxation time, remain finite as one moves towards the critical point but their slopes diverge at the critical point and behave like  $(\frac{\mu}{T}^*-\frac{\mu}{T})^{-\theta}$ \cite{ Ebrahim:2017gvk, Ebrahim:2018uky, Ebrahim:2020qif}. Now, we like to check the behavior of EoP near this critical point for two different regimes discussed earlier.

In order to find the slope of EoP near critical point, we start with 
\begin{align}
 \frac{d E_p}{d(\frac{\mu}{T})}=\frac{d E_p}{d\xi}\frac{d\xi}{d(\frac{\mu}{T})},
\end{align}
where $\frac{d\xi}{d(\frac{\mu}{T})}$ can be easily found from \eqref{QzH}. $\frac{d E_p}{d\xi}$ is obtained by taking the derivative of \eqref{ahmad} for low temperature and of \eqref{EoPP3} for intermediate temperature. Therefore, after expanding in power of $(\frac{\mu}{T}^*-\frac{\mu}{T})$, we have \\
\\
{\it{\textbf{Low temperature:}}}\\
\begin{align}
\begin{split}
\frac{d E_p}{d(\frac{\mu}{T})}=\frac{L^2\pi^\frac{7}{2} l(l'+l)T^4}{2^\frac{1}{4} a_1^3}
(a_1-a_3) \left(\frac{\mu}{T}^*-\frac{\mu}{T}\right)^{-\h}+
{\cal{O}}\left((\m^*-\frac{\mu }{T})^0\right)
\end{split}
\end{align}
\\
{\it{\textbf{Intermediate temperature:}}}\\
\begin{align}\label{high EoP slope}
\frac{d E_p}{d(\frac{\mu}{T})}=C \left(\m^*-\frac{\mu }{T}\right)^{-\frac{1}{2}}+ \mathcal{O}\left((\m^*-\frac{\mu }{T})^0\right),
\end{align}
where
\begin{align}
\begin{split}
C&=\frac{L^2}{2}\vast\lbrace\sum\limits_{n=2}^{\infty}\frac{-\pi \Gamma\left(n+\frac{1}{2}\right)T^2}{2^\frac{5}{4}(n-1)n\Gamma(n+1)}\cr
&\times\left[nF_1\left(n-1;\frac{-1}{2},\frac{1}{2};n;-2,-3\right)+3(n-1)F_1\left(n;\frac{-1}{2},\frac{3}{2};n+1;-2,-3\right)-F_1\left(n;\frac{1}{2},\frac{1}{2};n+1;-2,-3\right)\right]\cr
&+\frac{\pi^\frac{3}{2}T^2}{12\times 2^\frac{1}{4}}\left[13+9\sqrt{3}+12F_1\left(1;\frac{1}{2},\frac{1}{2};2;-2,-3\right)-9F_1\left(2;\frac{1}{2},\frac{3}{2};3;-2,-3\right)-9F_1\left(2;\frac{3}{2},\frac{1}{2};3;-2,-3\right)\right]\cr
&-\frac{\pi^\frac{7}{2}l'^2T^4}{12\times 2^\frac{1}{4}a_1^3}\left(19a_1-18a_2-10a_3\right) \vast\rbrace.
\end{split}
\end{align}
As a result, our simple calculation shows that the slope of EoP diverges near the critical point like other observables and dynamical critical exponent is equal to 0.5.


\begin{thebibliography}{99}

%\cite{Bhattacharyya:2019tsi}
\bibitem{Bhattacharyya:2019tsi}
A.~Bhattacharyya, A.~Jahn, T.~Takayanagi and K.~Umemoto,
``Entanglement of Purification in Many Body Systems and Symmetry Breaking,''
Phys. Rev. Lett. \textbf{122} (2019) no.20, 201601
%doi:10.1103/PhysRevLett.122.201601
[arXiv:1902.02369 [hep-th]].
%22 citations counted in INSPIRE as of 05 Dec 2020

   % \cite{arXiv:quant-ph/0202044v3}
  \bibitem{arXiv:quant-ph/0202044v3}
  B.~M.~Terhal, M.~Horodecki, D.~W.~Leung and D.~P.~DiVincenzo,
  ``The entanglement of purification," 
  J.~Math.Phys. {\bf 43}, 4286 (2002)
  % 	doi:10.1063/1.1498001
  [arXiv:quant-ph/0202044v3].
  
  %\cite{Maldacena:1997re}
\bibitem{Maldacena:1997re}
J.~M.~Maldacena,
``The Large N limit of superconformal field theories and supergravity,''
Int. J. Theor. Phys. \textbf{38} (1999), 1113-1133
%doi:10.1023/A:1026654312961
[arXiv:hep-th/9711200 [hep-th]].
%16244 citations counted in INSPIRE as of 05 Dec 2020

%\cite{CasalderreySolana:2011us}
\bibitem{CasalderreySolana:2011us}
  J.~Casalderrey-Solana, H.~Liu, D.~Mateos, K.~Rajagopal and U.~A.~Wiedemann,
  ``Gauge/String Duality, Hot QCD and Heavy Ion Collisions,''
  book:Gauge/String Duality, Hot QCD and Heavy Ion Collisions. Cambridge, UK: Cambridge University Press, 2014
 5 doi:10.1017/CBO9781139136747
  [arXiv:1101.0618 [hep-th]].
  %%CITATION = doi:10.1017/CBO9781139136747;%%
  %699 citations counted in INSPIRE as of 05 Jan 2021

%\cite{Hartnoll:2009sz}
\bibitem{Hartnoll:2009sz}
S.~A.~Hartnoll,
``Lectures on holographic methods for condensed matter physics,''
Class. Quant. Grav. \textbf{26} (2009), 224002
%doi:10.1088/0264-9381/26/22/224002
[arXiv:0903.3246 [hep-th]].
%1320 citations counted in INSPIRE as of 05 Jan 2021

%\cite{Ryu:2006bv}
\bibitem{Ryu:2006bv}
S.~Ryu and T.~Takayanagi,
``Holographic derivation of entanglement entropy from AdS/CFT,''
Phys. Rev. Lett. \textbf{96} (2006), 181602
%doi:10.1103/PhysRevLett.96.181602
[arXiv:hep-th/0603001 [hep-th]].
%2493 citations counted in INSPIRE as of 05 Dec 2020

%\cite{Takayanagi:2017knl}
\bibitem{Takayanagi:2017knl}
  T.~Takayanagi and K.~Umemoto,
  ``Entanglement of purification through holographic duality,''
  Nature Phys.\  {\bf 14}, no. 6, 573 (2018)
  %doi:10.1038/s41567-018-0075-2
  [arXiv:1708.09393 [hep-th]].
  %%CITATION = doi:10.1038/s41567-018-0075-2;%%
  %89 citations counted in INSPIRE as of 05 Dec 2020
  
%\cite{Nguyen:2017yqw}
\bibitem{Nguyen:2017yqw}
  P.~Nguyen, T.~Devakul, M.~G.~Halbasch, M.~P.~Zaletel and B.~Swingle,
  ``Entanglement of purification: from spin chains to holography,''
  JHEP {\bf 1801}, 098 (2018)
 % doi:10.1007/JHEP01(2018)098
  [arXiv:1709.07424 [hep-th]].
  %%CITATION = doi:10.1007/JHEP01(2018)098;%%
  %68 citations counted in INSPIRE as of 05 Dec 2020
  
%\cite{Kudler-Flam:2018qjo}
\bibitem{Kudler-Flam:2018qjo}
J.~Kudler-Flam and S.~Ryu,
``Entanglement negativity and minimal entanglement wedge cross sections in holographic theories,''
Phys. Rev. D \textbf{99} (2019) no.10, 106014
%doi:10.1103/PhysRevD.99.106014
[arXiv:1808.00446 [hep-th]].
%60 citations counted in INSPIRE as of 05 Dec 2020

%\cite{Tamaoka:2018ned}
\bibitem{Tamaoka:2018ned}
K.~Tamaoka,
``Entanglement Wedge Cross Section from the Dual Density Matrix,''
Phys. Rev. Lett. \textbf{122} (2019) no.14, 141601
%doi:10.1103/PhysRevLett.122.141601
[arXiv:1809.09109 [hep-th]].
%57 citations counted in INSPIRE as of 05 Dec 2020

%\cite{Agon:2018lwq}
\bibitem{Agon:2018lwq}
C.~A.~Ag\'on, J.~De Boer and J.~F.~Pedraza,
``Geometric Aspects of Holographic Bit Threads,''
JHEP \textbf{05} (2019), 075
%doi:10.1007/JHEP05(2019)075
[arXiv:1811.08879 [hep-th]].
%37 citations counted in INSPIRE as of 05 Dec 2020

%\cite{Dutta:2019gen}
\bibitem{Dutta:2019gen}
S.~Dutta and T.~Faulkner,
``A canonical purification for the entanglement wedge cross-section,''
[arXiv:1905.00577 [hep-th]].
%65 citations counted in INSPIRE as of 05 Dec 2020
  
  %\cite{Gubser:1998jb}
\bibitem{Gubser:1998jb}
S.~S.~Gubser,
``Thermodynamics of spinning D3-branes,''
Nucl. Phys. B \textbf{551} (1999), 667-684
%doi:10.1016/S0550-3213(99)00194-7
[arXiv:hep-th/9810225 [hep-th]]; 
K.~Behrndt, M.~Cvetic and W.~A.~Sabra,
``Nonextreme black holes of five-dimensional N=2 AdS supergravity,''
Nucl. Phys. B \textbf{553} (1999), 317-332
%doi:10.1016/S0550-3213(99)00243-6
[arXiv:hep-th/9810227 [hep-th]]; 
P.~Kraus, F.~Larsen and S.~P.~Trivedi,
``The Coulomb branch of gauge theory from rotating branes,''
JHEP \textbf{03} (1999), 003
%doi:10.1088/1126-6708/1999/03/003
[arXiv:hep-th/9811120 [hep-th]]; 
R.~G.~Cai and K.~S.~Soh,
%``Critical behavior in the rotating D-branes,''
Mod. Phys. Lett. A \textbf{14} (1999), 1895-1908
doi:10.1142/S0217732399001966
[arXiv:hep-th/9812121 [hep-th]];
M.~Cvetic and S.~S.~Gubser,
``Phases of R charged black holes, spinning branes and strongly coupled gauge theories,''
JHEP \textbf{04} (1999), 024
%doi:10.1088/1126-6708/1999/04/024
[arXiv:hep-th/9902195 [hep-th]]; 
M.~Cvetic and S.~S.~Gubser,
``Thermodynamic stability and phases of general spinning branes,''
JHEP \textbf{07} (1999), 010
%doi:10.1088/1126-6708/1999/07/010
[arXiv:hep-th/9903132 [hep-th]].
%125 citations counted in INSPIRE as of 05 Jan 2021
  
%\cite{DeWolfe:2011ts}
\bibitem{DeWolfe:2011ts}
O.~DeWolfe, S.~S.~Gubser and C.~Rosen,
``Dynamic critical phenomena at a holographic critical point,''
Phys. Rev. D \textbf{84} (2011), 126014
%doi:10.1103/PhysRevD.84.126014
[arXiv:1108.2029 [hep-th]].
%76 citations counted in INSPIRE as of 05 Dec 2020
  
%\cite{Finazzo:2016psx}
\bibitem{Finazzo:2016psx}
S.~I.~Finazzo, R.~Rougemont, M.~Zaniboni, R.~Critelli and J.~Noronha,
``Critical behavior of non-hydrodynamic quasinormal modes in a strongly coupled plasma,''
JHEP \textbf{01} (2017), 137
%doi:10.1007/JHEP01(2017)137
[arXiv:1610.01519 [hep-th]].
%22 citations counted in INSPIRE as of 05 Dec 2020
  
  %\cite{Ebrahim:2020qif}
\bibitem{Ebrahim:2020qif}
H.~Ebrahim and G.~M.~Nafisi,
``Holographic Mutual Information and Critical Exponents of the Strongly Coupled Plasma,''
Phys. Rev. D \textbf{102} (2020) no.10, 106007
%doi:10.1103/PhysRevD.102.106007
[arXiv:2002.09993 [hep-th]].
%4 citations counted in INSPIRE as of 05 Dec 2020
  
 %\cite{BabaeiVelni:2019pkw}
\bibitem{BabaeiVelni:2019pkw}
K.~Babaei Velni, M.~R.~Mohammadi Mozaffar and M.~H.~Vahidinia,
``Some Aspects of Entanglement Wedge Cross-Section,''
JHEP \textbf{05} (2019), 200
%doi:10.1007/JHEP05(2019)200
[arXiv:1903.08490 [hep-th]].
%27 citations counted in INSPIRE as of 05 Dec 2020 
  
 %\cite{Ebrahim:2017gvk}
\bibitem{Ebrahim:2017gvk}
H.~Ebrahim and M.~Ali-Akbari,
``Dynamically probing strongly-coupled field theories with critical point,''
Phys. Lett. B \textbf{783} (2018), 43-50
%doi:10.1016/j.physletb.2018.06.048
[arXiv:1712.08777 [hep-th]].
%5 citations counted in INSPIRE as of 05 Dec 2020
  
%\cite{Ebrahim:2018uky}
\bibitem{Ebrahim:2018uky}
H.~Ebrahim, M.~Asadi and M.~Ali-Akbari,
``Evolution of Holographic Complexity Near Critical Point,''
JHEP \textbf{09} (2019), 023
%doi:10.1007/JHEP09(2019)023
[arXiv:1811.12002 [hep-th]].
%4 citations counted in INSPIRE as of 05 Dec 2020

%\cite{Amrahi:2020jqg}
\bibitem{Amrahi:2020jqg}
B.~Amrahi, M.~Ali-Akbari and M.~Asadi,
``Holographic Entanglement of Purification near a Critical Point,''
Eur. Phys. J. C \textbf{80} (2020) no.12, 1152
[arXiv:2004.02856 [hep-th]].
%1 citations counted in INSPIRE as of 17 Dec 2020

\end{thebibliography}
\end{document}